\documentclass{acmart}
\usepackage{subcaption}
\usepackage{graphicx}
\usepackage{multirow}
\usepackage[export]{adjustbox}
\usepackage{bbm}
\usepackage{makecell}
\usepackage[normalem]{ulem}
\AtBeginDocument{%
  \providecommand\BibTeX{{%
    \normalfont B\kern-0.5em{\scshape i\kern-0.25em b}\kern-0.8em\TeX}}}

\setcopyright{none}
\renewcommand\footnotetextcopyrightpermission[1]{}
\copyrightyear{2022}
\acmYear{2022}
\acmDOI{XXXXXXX.XXXXXXX}

\acmConference[FashionXRecSys '22]{Fourth Workshop on Recommender Systems in Fashion}{September 18--23,
  2022}{Seattle, WA, USA}
\acmBooktitle{FashionXRecSys'22: Workshop on Recommender Systems in Fashion,
 September 23, 2022, Seattle, WA} 
\acmPrice{15.00}
\acmISBN{978-1-4503-XXXX-X/18/06}

\begin{document}

\title{Reusable Self-Attention-based Recommender System for Fashion}

\author{Marjan Celikik}
\email{marjan.celikik@zalando.de}
\author{Jacek Wasilewski}
\email{jacek.wasilewski@zalando.de}
\author{Sahar Mbarek}
\email{sahar.mbarek@zalando.de}
\author{Pablo Celayes}
\email{pablo.celayes@zalando.de}
\author{Pierre Gagliardi}
\email{pierre.gagliardi@zalando.de}
\author{Duy Pham}
\email{duy.pham@zalando.de}
\author{Nour Karessli}
\email{nour.karessli@zalando.de}
\author{Ana Peleteiro Ramallo}
\email{ana.peleteiro.ramallo@zalando.de}
\affiliation{
  \institution{Zalando SE}
  \city{Berlin}
  \country{Germany}
}

\renewcommand{\shortauthors}{Celikik, et al.}

\newcommand{\Autoref}[1]{%
  \begingroup%
  \def\chapterautorefname{Chapter}%
  \def\sectionautorefname{Section}%
  \def\subsectionautorefname{Subsection}%
  \def\figureautorefname{Figure}%
  \def\tableautorefname{Table}%
  \autoref{#1}%
  \endgroup%
}

\begin{abstract}
A large number of empirical studies on applying self-attention models in the domain of recommender systems are based on offline evaluation and metrics computed on standardized datasets, without insights on how these models perform in real life scenarios. Moreover, many of them do not consider information such as item and customer metadata, although deep-learning recommenders live up to their full potential only when numerous features of heterogeneous types are included. Also, typically recommendation models are designed to serve well only a single use case, which increases modeling complexity and maintenance costs, and may lead to inconsistent customer experience. In this work, we present a reusable Attention-based Fashion Recommendation Algorithm (AFRA), that utilizes various interaction types with different fashion entities such as items (e.g., shirt), outfits and influencers, and their heterogeneous features. Moreover, we leverage temporal and contextual information to address both short and long-term customer preferences. We show its effectiveness on outfit recommendation use cases, in particular: 1) personalized ranked feed; 2) outfit recommendations by style; 3) similar item recommendation and 4) in-session recommendations inspired by most recent customer actions. We present both offline and online experimental results demonstrating substantial improvements in customer retention and engagement.

\end{abstract}

\ccsdesc[500]{Computing methodologies~Neural networks}
\ccsdesc[500]{Information systems~Recommender systems}

\keywords{Recommendation Systems, Transformers, Fashion Industry}

\maketitle

\section{INTRODUCTION}

\begin{figure*}[t]
\centering
\begin{minipage}{0.43\linewidth}
\centering
\begin{subfigure}{\textwidth}
\includegraphics[width=\textwidth]{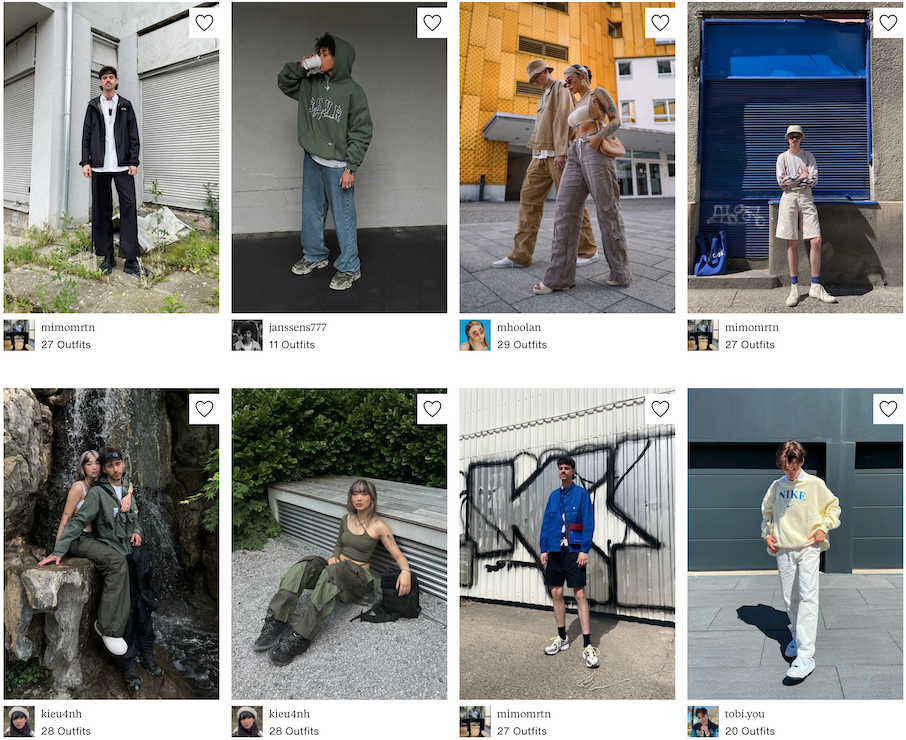}
\caption{"Get the Look" personalized feed (catalog) of ranked outfits.}
\label{fig:usecase-gtl}
\end{subfigure}
\end{minipage}%
\hfill
\begin{minipage}{0.53\linewidth}
\centering
\begin{subfigure}{\textwidth}
\includegraphics[width=\textwidth]{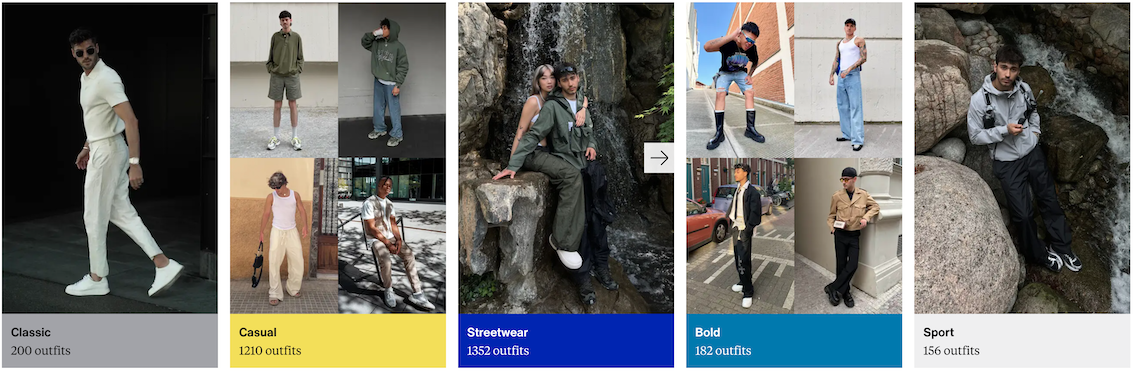}
\caption{"Style preview" carousel with personalized outfits selection in different styles.}
\label{fig:usecase-style}
\end{subfigure}

\begin{subfigure}{\textwidth}
\includegraphics[width=\textwidth]{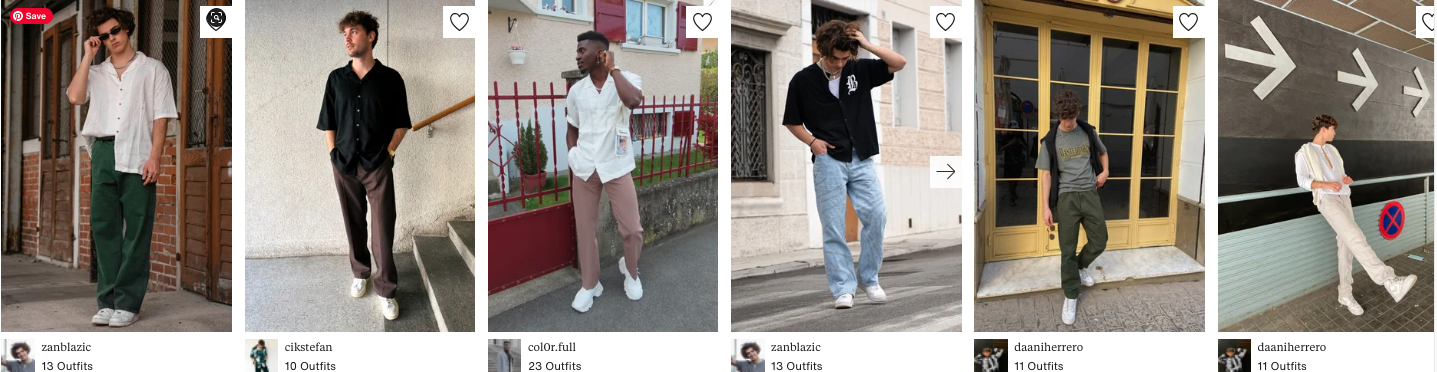}
\caption{"Inspired by you" carousel with in-session outfit recommendations based on customer's recent interactions.}
\label{fig:usecase-inspired}
\end{subfigure}
\end{minipage}

\begin{subfigure}{0.6\textwidth}
\includegraphics[width=0.19\textwidth,valign=t]{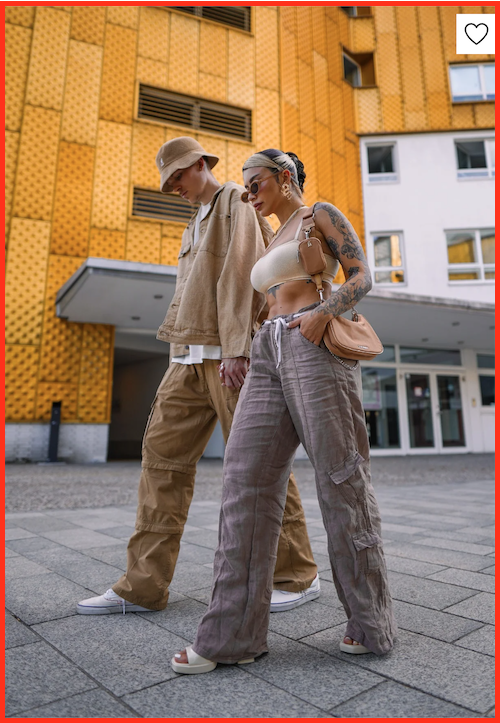}
\includegraphics[width=0.79\textwidth,valign=t]{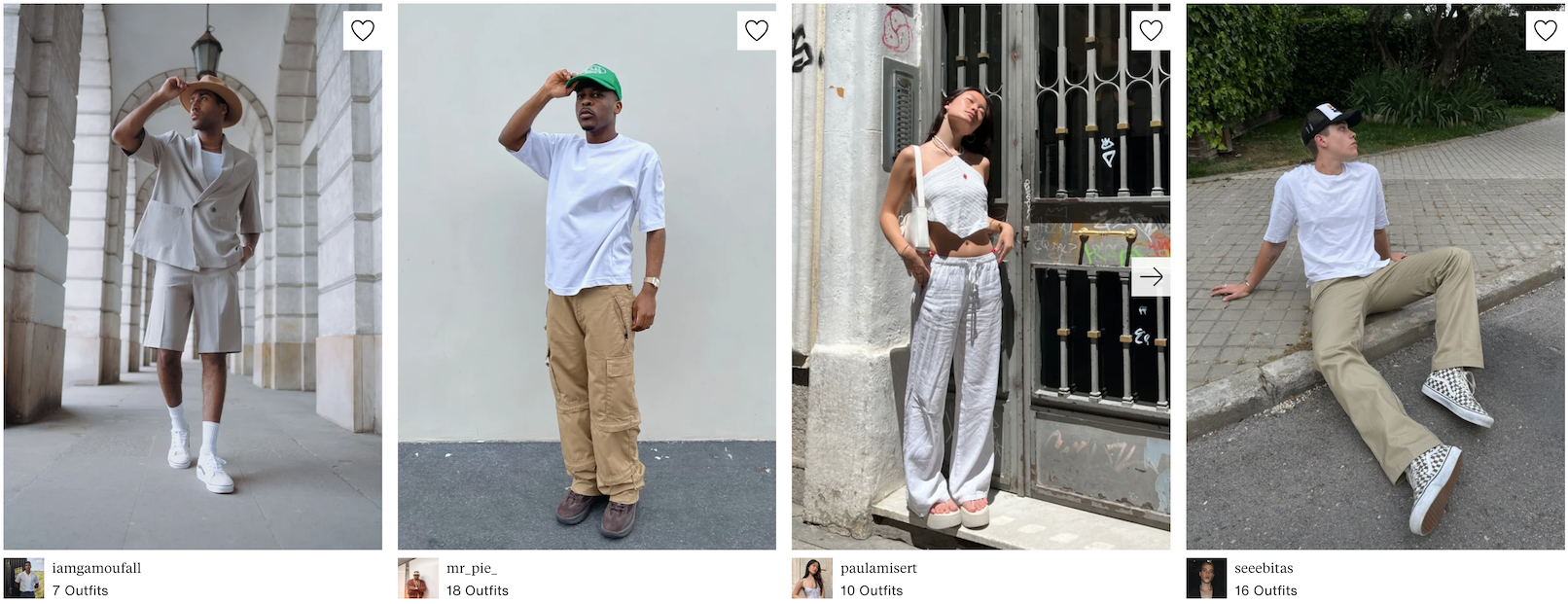}
\caption{"You might also like" carousel with personalized outfit recommendations (right) related to an anchor outfit (left).}
\label{fig:usecase-youmight}
\end{subfigure}

\caption{Different outfit recommendation and ranking use cases.}
\label{fig:usecases}
\end{figure*}

Fashion contributes to our everyday life in reflecting personality, culture, and style. With the vast choice of items available in e-commerce, it has become increasingly difficult for customers to find relevant content, combine it and match with a specific style. Finding inspiration, such as for example influencers, has become crucial to support customers in discovering inspirational and fresh outfits tuned to their taste.   

Zalando is a leading European online platform for fashion and lifestyle, and one of its main goals is to provide fashion inspiration and support customers in their fashion discovery journey. We do this through different experiences, such as for example creator outfit inspirational content, in touch points such as: a) "Get the Look" (GTL), a personalized feed of all available influencer outfits (\Autoref{fig:usecase-gtl});
b) "Style" carousel, providing personalized outfit recommendations in different styles (e.g. classic, casual, streetwear; \Autoref{fig:usecase-style}); c) "Inspired by you" carousel, providing real-time in-session recommendation based on the customer’s recent interactions (\Autoref{fig:usecase-inspired}); and d) "You might also like" carousel, providing personalized recommendations related to the currently viewed outfit (\Autoref{fig:usecase-youmight}).

A complexity that arises when building recommendation models for inspirational content in the e-commerce domain is the availability of input signals. Interaction data with non-product fashion entities, such as influencer outfits, is usually low, which makes the data extremely sparse when compared to interactions with individual items (e.g., a shirt). In this work, we investigate how interactions with other "high-traffic" fashion entities could be used and improve "low-traffic" entities such as outfit recommendations. 

In e-commerce, often different but related customer experiences are served by completely different recommender systems, for example either to steer more to in-session recommendations or to capture longer-term customer preferences. This can significantly increase not only modeling complexity and maintenance costs of those different systems, but also lead to inconsistent customer experience. In contrast to the common belief that different recommenders are needed for different use cases \cite{netflixdl}, we show that a single Transformer-based recommender system can be trained on diverse types of interactions coming from various sources and successfully re-used across many related use cases, such as session-based recommendation for short-term interests, personalized ranking based on long-term user preferences and even provide item-related or similar recommendations. 

Moreover, in e-commerce, many users are either new or do not login when navigating, thus it is important to use a model that can provide meaningful recommendations to new, cold-start users. The model we propose can serve "partial" cold-start users (users that have not interacted with outfits but have interacted with the platform) by utilizing other types of interactions in the session, as well as "full" cold-start users (without any platform activity) by utilizing contextual information about the user and the use case, such as premise (site or product), device, country, language, etc.

In recent years sequential recommendation approaches have gained significant traction \cite{huifang}. These model customers as sequences of interactions to better infer context and customer preferences that change over time. In addition, representing customers as sequences of actions at least partially simplifies the feature engineering and modeling efforts. Sequential recommendation approaches often employ NLP techniques such as Word2Vec \cite{mikolov}, RNNs \cite{hidasi2015sessionbased, rnns} and attention mechanisms \cite{sasrec, bert4rec, alibabapog}. Transformer-based models are the most promising sequential and session-based models to date due to their ability to model long-range sequences and their scalability thanks to efficient parallel training \cite{trm4rec, sasrec, bert4rec}. However, the majority of previous studies are based on offline evaluation and metrics computed on open datasets, which may not be directly comparable with industrial datasets and online results. Moreover, many of them do not consider side information such as categorical inputs to represent items or customers, although it is well known that deep learning recommender systems live up to their full potential only when numerous features of heterogeneous types are used \cite{netflixdl}. Due to these shortcomings, even if relevant, most of the research work on self-attention models for recommendations does not explore the potential effectiveness in real-world industry applications. In this work, we contribute to bridge this gap by presenting online experimental results of using a self-attention-based recommender system that can utilize contextual information about the customer and the use case. We showcase that our personalized outfit recommendation model can improve engagement and customer retention in different use cases throughout the customer journey. 

In summary, the main contributions of this work are as follows: (1) a reusable transformer-based recommender system for fashion recommendations that utilizes diverse types of interactions on various fashion entities. It is able to provide session-based recommendations as well as take into consideration long-term user preferences and contextual information, to both recurring and cold-start users; (2) extensive A/B testing that shows that this approach can be successfully applied to different recommendation use cases including personalized ranked feed, outfit recommendations by style, similar item recommendation, and in-session recommendations inspired by most recent customer activities; (3) extensive offline experiments and evaluation considering different ranking losses and metrics where we show how our approach substantially increases customer engagement and retention. The rest of the paper is structured as follows: in \Autoref{related-work} we discuss related work; \Autoref{algorithm} describes the proposed approach; we present offline evaluation in \Autoref{offline-evaluation} and online experiment results in \Autoref{online-evaluation}; finally, we provide discussion of conclusions and future work in \Autoref{conclusions}.

\section{RELATED WORK}
\label{related-work}
The main objective of recommender systems is to recommend relevant content tailored to the customer’s preferences and intent. Fashion e-commerce has heavily invested in developing recommender systems for different use cases to aid online shopping experience \cite{Chakraborty2021,Deldjoo2022} including recommending relevant items \cite{Deng2018,Cardoso2018,Zhou2018}, complete outfits \cite{Lin2019,Celikik2020,Denk2020}, and size recommendation \cite{sembium2018,Lasserre2020}.

Transformer-based recommendation systems started with the SASRec algorithm presented in \cite{sasrec}, where, similarly to GPT \cite{gpt}, a Transformer encoder with masked attention for causal language modeling (CLM) was used to predict the next item in the sequence of user interaction. A similar work in \cite{ssept} builds on this idea by concatenating user embeddings to each learned item embedding from the same sequence to add contextual user information. BERT4Rec \cite{bert4rec} is another work that employs the Transformer encoder trained with the masked language modeling (MLM) approach, while masking only the last item of the sequence during inference to avoid leaks of future information.

There are a few works in the literature that use complex hierarchical architectures or mix together different network architectures. The SDM algorithm introduced in \cite{sdm} is fusing LSTMs and self-attention networks to obtain a final user behavior vector. \cite{seqhierarchicalattention} proposes a similar approach by using a two-layer hierarchical attention-based network while \cite{graphselfattention} employs a graph and a self-attention network. In all of these approaches, no evidence is provided as to why the complex architecture is needed. Moreover, these methods are only compared to baselines that are either not based on self-attention or not based on deep learning methods. Hence, it is unclear whether the improvement comes from the choice of the architecture. In our case, we present an effective and simple approach that is able to capture both short-term and long-term user interests.

Most of the published studies do not consider side information for representing items or users and instead work with user-item interaction data only. The few works that include side information usually consider only a few item features such as category and brand \cite{sdm}. The BST algorithm from Alibaba \cite{alibabapog} considers richer item features and user profiles as well as contextual features. The differences to our approach are twofold: first, the contextual features are not part of the self-attention mechanism but rather concatenated with the output of the Transformer, and second, the use of a positional encoding for the entire sequence instead of representing different sessions separately.

There are only a few lines of work that we are aware of that consider (re)using models for multiple recommendation use cases. Parallel to our work, the work in \cite{rnnlong} shows that RNNs can perform well in both, short and long-term recommendation tasks when certain improvements are applied and even over-perform more complex hierarchical models. The work \cite{multigraph} presents a multi-graph structured multi-use case recommendation solution which encapsulates interaction data across various use cases and demonstrates increase in CTR and video views per user. Another work in \cite{knowledgetransferrecsys} surveys the benefits of pre-training and knowledge transfers in recommender systems in order to alleviate the data sparsity problem.
\section{ALGORITHM}
\label{algorithm}

\begin{figure*}[t]
  \centering
  \includegraphics[width=\linewidth]{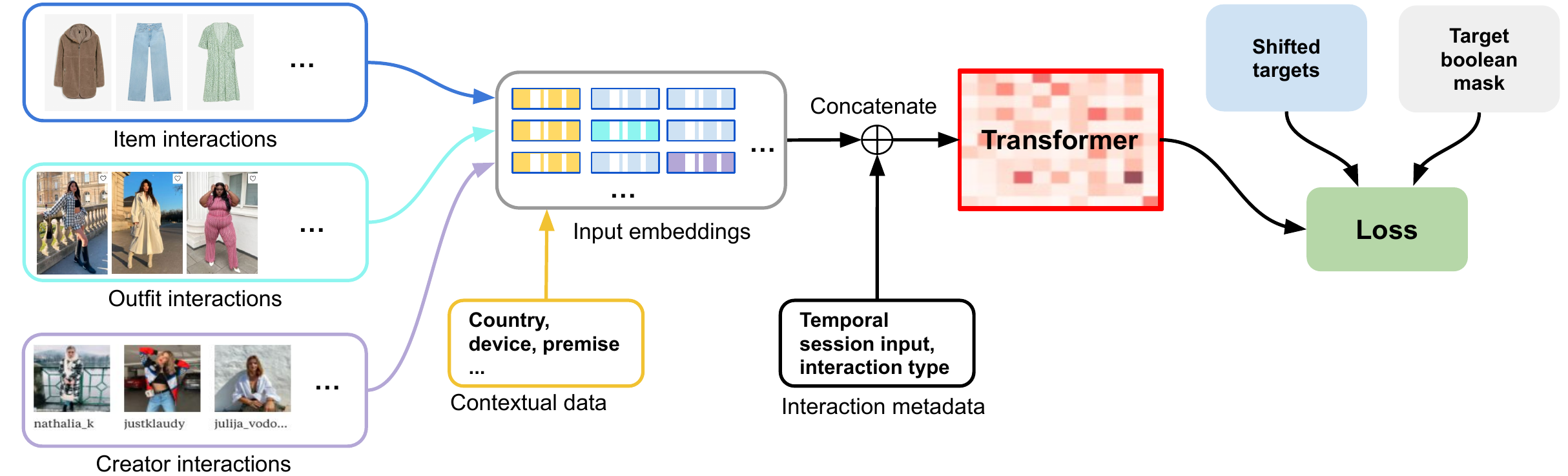}
  \caption{Overview of the model trained on interactions coming from different fashion entities. Interactions are converted into embeddings of the same structure in terms of feature representation. Contextual features corresponding to the use case such as premise (site) and device are added to input sequences, where temporal session data and interaction metadata is concatenated directly to the learned embeddings. Masking is used in order to consider target outputs only of certain entity types (e.g. outfits).}
  \label{fig:model}
\end{figure*}

\subsection{Problem Formulation}
Let $\mathcal{U}=\{u_1, u_2, ..., u_{|\mathcal{U}|}\}$ denote the set of users and we denote a set of items by $\mathcal{V}=\{v_1, v_2, ..., v_{|\mathcal{V}|}\}$, we represent the interaction sequence as $\mathcal{S}_u=\{v_1^{(u)}, ..., v_t^{(u)}, ..., v_{n_u}^{(u)}\}$ with interactions ordered chronologically for the user $u \in \mathcal{U}$, where $v_t^{(u)} \in \mathcal{V}$ is the item that the user $u$ has interacted with at time $t$, and $n_u$ is the length of interaction sequence for user $u$. Let $\mathcal{C}_u$ be the contextual information about user $u$, such as country, device, premise and date. Given the interaction history $\mathcal{S}_u$ and context $\mathcal{C}_u$, sequential recommender system aims to predict the item that user $u$ is most likely to interact with at time $n_u + 1$, i.e. we would like to find an item $i$ so that the probability $p \left(v_{n_u + 1}^{(u)}=i | \mathcal{S}_u, \mathcal{C}_u \right)$ is maximized.

\subsection{Model Architecture}

\begin{figure*}[t]
    \centering
    \includegraphics[width=\textwidth]{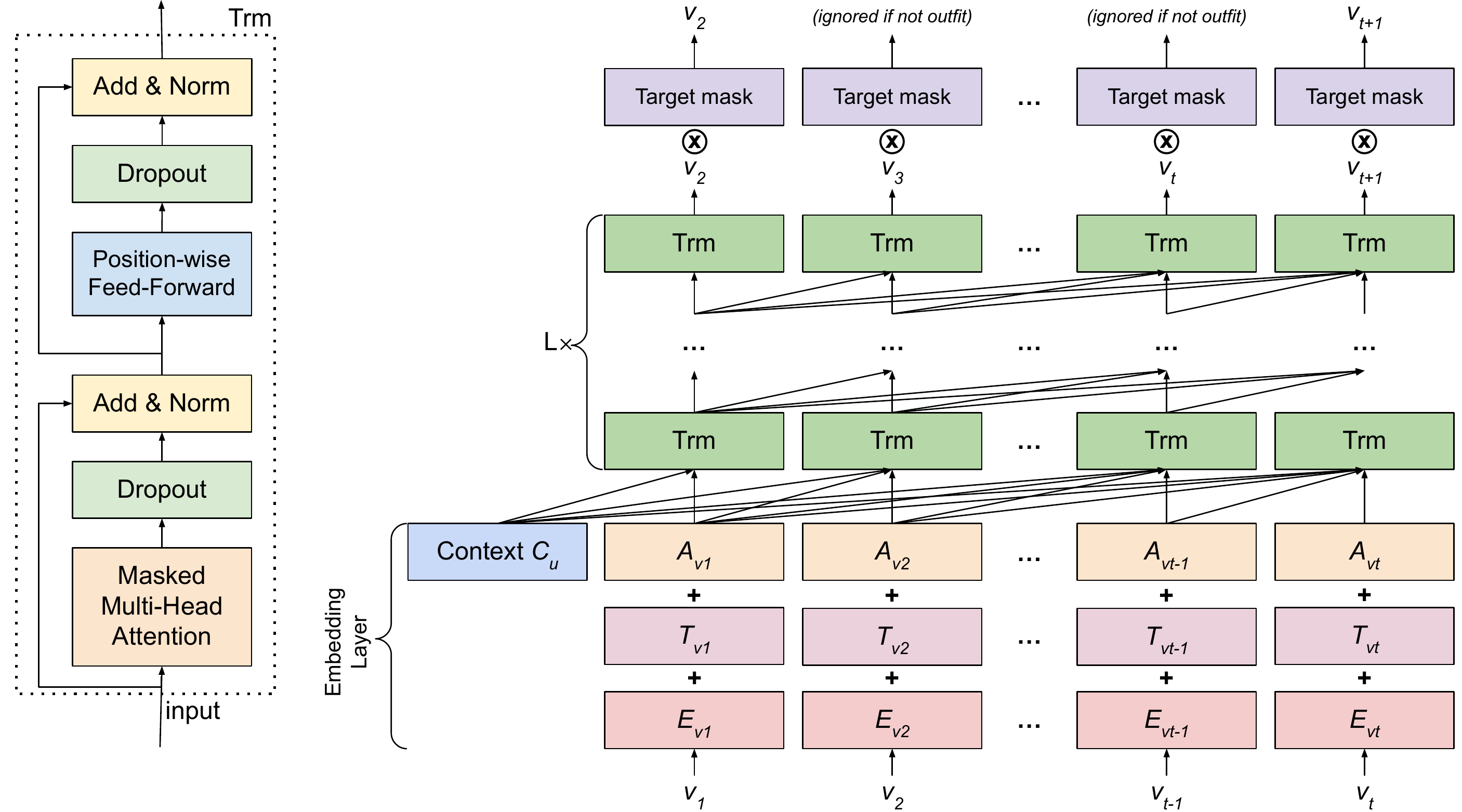}
    \caption{Illustration of a single Transformer encoder (left) and overview of our model architecture (right). Model inputs consist of contextual information and a sequence of: item embedding $E_{v_t}$, action embedding $A_{v_t}$ and session embedding $T_{v_t}$. Target mask filters predictions of entities other than the target ones -- in this example target mask is set for outfit entities.}
    \label{fig:transformer}
\end{figure*}

We use a standard Transformer encoder \cite{sasrec,attention,trm4rec} trained with the causal language model (CLM) approach as illustrated in \Autoref{fig:transformer}. A causal attention mask is provided to the self-attention mechanism so that each position can attend only to earlier positions in the sequence. Note that our approach is oblivious of the training logic and it can be trained with the masked language model (MLM) approach as well \cite{bert4rec,trm4rec}. Given an input sequence of length $t$ and a matrix of learnable input embeddings $V \in \mathbb{R}^{t \times d}$, a single layer of the Transformer model computes hidden representations $H^1=\text{Trm}(V) \in \mathbb{R}^{t \times d}$. By stacking multiple layers, we have $H^l=\text{Trm}(H^{l-1})$. For details we refer the reader to \cite{attention} and \cite{sasrec}. The final output of the last position is fed into a softmax layer over the item set. We employ categorical cross-entropy as a loss function, however, experiment with other losses as well.

Given an input sequence, the expected target output is a shifted version of the same sequence. In our setting, an item can refer to a fashion article, an outfit or an influencer. We train on every item in a sequence, but predict only those items that are relevant for the use case. To that end, we assign a boolean mask to each relevant item. The mask is set to 1 only if the next position in the sequence is associated with an item that should be predicted by the model as a valid recommendation. For example, if our model recommends outfits, then all inputs corresponding to valid outfits will have a mask of 1, for other entities the mask will be set to 0 (all use cases presented in \Autoref{fig:usecases} consider outfits, so in our experiments we focus on outfit predictions). The mask is passed to the loss function and the positions corresponding to zeroes do not contribute to the loss. In addition, items that are valid but not available (e.g. out of stock), will have an output mask set to 0 as well.

Contextual information about the use case and the customer are encoded as embeddings with the same dimensionality as the input item embeddings and are set as the first positions of the sequence so that every other position can attend and utilize this information when making predictions. Inputs such as location, market and language play an important role for cold-start customers that are new to the platform and do not have any interactions yet. For cold-start customers that are new to the outfit use cases but not new to the platform itself, the model can make recommendations based on interactions with other fashion entities, either historical or from the current session. Inputs such as premise and device help the model to hone in on the particular use case. \Autoref{fig:model} provides a summary of the modeling choices and the different sources of data used for training.

\subsection{Input Embeddings}

We represent each item $v_t$ (for brevity we omit the superscript) in a user sequence $\mathcal{S}_u$ as a concatenation of learned embeddings that correspond to its different features. The features are always encoded in the same order. Since each input position in the sequence must consist of the same embeddings, we pad the corresponding part of the input tensor with zeroes if the item does not have a certain feature. For example, an outfit might have an influencer (creator), however, a single article does not have one. An item $v_t$ will have the following representation:
\begin{equation*}
    E_{v_t} = E_{f_1}(v_t) \oplus E_{f_2}(v_t) \oplus ... \oplus E_{f_n}(v_t) 
\end{equation*}
where $E_{f_k}$ is the embedding matrix of a feature $f_k$ and $\oplus$ is the concatenation operator. Depending on the feature type and its cardinality, we represent features through learned embedding matrices or by using 1-hot encoded vectors. Note that if $v_t$ does not have the feature $f_k$, $E_{f_k}(v_t)$ is the 0-vector. Examples of categorical features that we represent as embedding matrices are brand, category, color and influencer, while features such as average price bucket and outfit style are represented by using 1-hot encoding. 

As there are clear hierarchical relationships between fashion entities, e.g. an outfit consists of articles and influencers can be represented by the set of outfits they have created, we employ a simple representation of more complex items by averaging the embeddings of the individual items. For example, given an outfit $o$, let $E_B$ be an embedding matrix for the brand feature. We represent the brand embedding of outfit $o$ as
\begin{equation*}
    E_B(o)=\sum_{a \in o} E_B(a)
\end{equation*}
where $a$ is an article in $o$. We represent individual articles in a sequence as outfits with a single article.

For the model to be able to utilize the interaction type of an item and learn to treat different interactions differently, we concatenate a one-hot encoding of the interaction type of the item with the item representation, i.e. $E_{v_t} \oplus A_{v_t}$, where $A_{v_t}$ is one-hot encoding of the $v_t$’s interaction type.

\subsection{Modeling Sessions for Long and Short-Term Interests}

In this section we present a simple but yet effective mechanism that allows our algorithm to utilize information from the current and past browsing sessions in order to model short in-session and long-term customer preferences. A user sequence $\mathcal{S}_u$ consists of different browsing sessions. A session is a list of consecutive interactions within a given time frame, for example a day, when the user has a clear shopping intent while their interests can change sharply when they start a new session, say a week later. Hence, modeling user sequences directly while ignoring this structure will affect performance negatively (as observed in our experiments).

We model sessions via introducing temporal inputs in the form of interaction recency defined as number of days passed since the action has taken place relative to the model training timestamp during training and the serving timestamp during serving. We discretize recency and consider only the integer part of the timestamp. Each temporal input is assigned its own interaction recency that is concatenated with the rest of the item embeddings. For simplicity, we consider user activity during a single day as a single session. Hence, interactions within the same session will have the same recency. For example, if the next prediction is in-session then the recency of the last action will be 0. However, the model is able to attend to previous actions as well and use the attention mechanism to select actions that are relevant for the prediction.

Note that with CLM training, the model does not have bidirectional access to positions in the sequence, hence it has no information about when the next action has taken place. However, this information is needed for a correct prediction. Namely, the next action may happen on the same day or a week after the previous action. To account for this, in addition to recency, we include the time gap between two actions and make this information accessible to the previous action as one of its inputs. At inference time, the time gap is always 0. Hence, the final input representation of a single interaction to the model is 
\begin{equation*}
E_{v_t} \oplus T_{v_t} \oplus A_{v_t}
\end{equation*}
where $E_{v_t}$ is the item embedding, $A_{v_t}$ is the action embedding for item $v_t$, and $T_{v_t}=r_{v_t} \oplus t_{v_t}$ is the session embedding of item $v_t$, $r_{v_t}$ is the learned embedding of the discretized recency and $t_{v_t}$ is the time gap between interactions with item $v_t$ and item $v_{t+1}$.

\section{OFFLINE EVALUATION}
\label{offline-evaluation}

\subsection{Dataset}

Our data consists of a sample of 60 days of user interactions on the platform. We perform a time-based split, where we train on the first 59 days of data and evaluate on the last day. We aggregate the interaction per user into sequences. During the evaluation, we feed only those interactions as inputs that took place before the split timestamp (if any). We filter out sequences without outfit interactions. Our training data contains roughly 6.6m sequences on 9.9k distinct outfits created by 677 unique creators. The test data contains roughly 172k interactions on 9.4k distinct outfits. The average outfit length was 4.6 articles. The total number of distinct articles in the dataset is 1.6m. Roughly 23\% of the customers are completely new to the platform, 31\% new to outfits but have interactions with articles on the platform and 46\% of the customers have interactions with outfits. To confirm the consistency of our results, we set fixed random seeds, train and evaluate on 3 different datasets (collected with respect to 3 different last days) and report the average performance.

\subsection{Experimental Setting}

We evaluate our new Attention-based Fashion Recommendation Algorithm (AFRA) against a set of existing recommendation algorithms some of which have been powering outfit recommendation use cases presented in \Autoref{fig:usecases}. In the following we present each of the compared algorithms.

\noindent\textbf{Neural LTR}: powers the "Get the Look" personalized feed and the "Style" carousel use cases. It is a neural Learning-To-Rank approach implemented by using the TensorFlow-Ranking \cite{TensorflowRankingKDD2019} library to rank outfits based on outfit and user inputs. The outfit inputs consist of learned embeddings of categorical features such as brand, influencer, color and the market of the outfit. The user features are represented by the top-3 favorites brand, colors, influencers, market as well as historical normalized frequencies of how often the customer interacted with certain content types. A feed-forward network is used as the model architecture, with 2 dense layers with 128 and 64 layers units each. The output is passed through a $tanh$ activation function. A pairwise logistic loss is used for training. For the "Style" carousel use case, the output is further filtered by style.

\noindent\textbf{IB-CNN-kNN}: is a kNN-based recommendation algorithm powering the "Inspired by you" carousel. Outfit recommendations are retrieved based on cosine similarity between item embeddings of previously interacted fashion articles and item embeddings of articles in the outfit. The similarity is defined as the average cosine similarity between the embeddings of the best matching article from the user's history, considering the last 10 article interactions in near real-time and the 200 most recent outfits on the platform. Each item is represented by latent embeddings computed by using a fine-tuned CNN as in \cite{fdna}.

\noindent\textbf{IB-CF-kNN}: is a kNN-based Collaborative-Filtering algorithm powering the "You might also like" carousel. We retrieve outfit recommendations based on cosine similarity between user-outfit vectors.
 
\noindent\textbf{AFRA}: AFRA has been implemented as described in \Autoref{algorithm} by using the CLM training approach, 2 Transformer layers each with 8 attention heads, dff layer with 1024 units, dropout of 0.1,  $d_{model}$ set to 128, batch size of 64 and learning rate set to 0.01. The algorithm was trained with 10 epochs. Each interaction sequence passed to the model has been truncated to the 100 most recent interactions. The model uses a set of article-based categorical features such as brand, color, material, fit, pattern; outfit based-features such as influencer and style, general interaction features such interaction type and interaction recency; and contextual features such as premise, market and device type. We employ two strategies to evaluate AFRA: \textbf{AFRA-RT} in which the customer actions are available to the algorithm in near real-time and \textbf{AFRA-Batch} in which only customer actions performed up to the previous day are available to the algorithm.

\noindent\textbf{SASRec}: SASRec \cite{sasrec} is a Transformer-based recommendation algorithm trained with the CLM approach. We apply the same hyper-parameters that we use in AFRA. We experiment with the authors' implementation of the algorithm. The model uses only embedding representation of user and item IDs as inputs. The model is trained on both outfit and article interactions, while predictions are made on outfits only.

It is worth noting that AFRA is able to make use of additional content, contextual or temporal features that other approaches reasonably cannot (CF, kNN, LTR) or do not (SASRec) use without introducing substantial changes in the algorithms.

\subsection{Offline Experiments}

In this section we evaluate different aspects of our proposed algorithm such as \emph{relevance}, \emph{diversity} and \emph{freshness}. We experiment by using different variants of AFRA by training on different types of fashion items. In addition, we evaluate the performance of AFRA by using other ranking loss functions that are commonly considered in the literature.

\subsubsection{Relevance} 

We run offline evaluation using a number of relevance metrics, including recall@k, precision@k, hitrate@k, nDCG@k and mAP@k according to their standard definition provided in \cite{offlineeval}. For each algorithm we calculate the metrics for the top $k$ results, where $k \in \{5, 15, 30\}$ which corresponds to the use cases that we consider, e.g. $k=15$ and $k=30$ corresponds to the first and the second page of results in a feed, while $k=5$ corresponds to a single carousel of recommendations. The recommendation task consists of predicting outfit clicks, regardless whether they are sequential (i.e. in-session), historical (i.e. when customers return from another day) or cold-start (when customers perform an action for the first time). Since all of the relevance metrics are strongly correlated with each other, for brevity we report only recall@k, which in turn correlates well with click-through-rate (CTR) \cite{hidasi2015sessionbased}.

\begin{table*}[t]
\centering
\caption{Relevance in terms of recall@k for all compared algorithms computed on new and on all customers. It can be observed that all AFRA variants perform the best in general for the given use cases. The label "outfits only" means that the model has been trained by using outfit interactions only.}
\label{table:offline}
\begin{tabular}{llccc}
\toprule
\textbf{Customer segment} & \textbf{Algorithm} & \textbf{Recall@5} & \textbf{Recall@15} & \textbf{Recall@30} \\
\midrule
\multirow{3}{8em}{\textbf{All  customers}} & \textbf{AFRA-RT} & \textbf{0.135} & \textbf{0.233} & \textbf{0.301} \\
 & \textbf{AFRA-Batch} & 0.081 & 0.156 & 0.224 \\
 & \textbf{AFRA-RT (outfits only)} & 0.093 & 0.158 & 0.214 \\
 & \textbf{AFRA-Batch (outfits only)} & 0.056 & 0.108 & 0.156 \\
 & \textbf{LTR} & 0.025 & 0.104 & 0.153 \\
 & \textbf{IB-CNN-kNN} & 0.056 & 0.078 & 0.110 \\
 & \textbf{IB-CF-kNN} & 0.052 & 0.073 & 0.102 \\
 & \textbf{SASRec} & 0.061 & 0.102 & 0.144 \\
 & \textbf{SASRec (outfits only)} & 0.051 & 0.075 & 0.099 \\
\midrule
\multirow{3}{8em}{\textbf{New customers (cold-start)}} & \textbf{AFRA-RT} & \textbf{0.082} & \textbf{0.164} & \textbf{0.232} \\
 & \textbf{AFRA-Batch} & 0.045 & 0.102 & 0.163 \\
 & \textbf{AFRA-RT (outfits only)} & 0.069 & 0.131 & 0.183 \\
 & \textbf{AFRA-Batch (outfits only)} & 0.045 & 0.093 & 0.134 \\
 & \textbf{LTR (w/ recency)} & 0.033 & 0.085 & 0.122 \\
 & \textbf{Popularity} & 0.030 & 0.052 & 0.091 \\
 & \textbf{IB-CNN-kNN} & 0.034 & 0.042 & 0.055 \\
 & \textbf{IB-CF-kNN} & 0.047 & 0.057 & 0.074 \\
 & \textbf{SASRec} & 0.041 & 0.074 & 0.101 \\
 & \textbf{SASRec (outfits only)} & 0.005 & 0.011 & 0.032 \\
\bottomrule
\end{tabular}
\end{table*}

\Autoref{table:offline} shows the recall@k for all the compared algorithms. The following main observations can be made. First, the two versions of AFRA that simulate real-time and batch scenarios perform better than all other algorithms, including SASRec, on all and on cold-start customers specifically. Second, training by using diverse item interaction data substantially improves the relevance compared to training on outfits only. This can be observed on both algorithms, AFRA and SASRec. Third, AFRA performs substantially better than SASRec even when trained on the much sparser outfit interactions only, thanks to using rich heterogeneous data and session encoding. Fourth, AFRA performs substantially better on cold-start customers compared to all other algorithms thanks to its ability to utilize contextual data. This is especially pronounced in the real-time use case where session data is available to the algorithm. Fifth, using real-time session data has a large impact on performance: in certain scenarios AFRA-RT achieves twice as high a recall compared to AFRA-Batch.

\begin{table*}[t]
\begin{minipage}{0.60\textwidth}
\centering
\caption{Change in AFRA’s recall@k when various loss functions are used compared to full cross entropy loss. In summary, the gain in training speed does not justify the drop in the offline evaluation metrics.}
\label{table:losses}
\begin{tabular}{lcc}
\toprule
\textbf{Loss} & \textbf{Recall@5 $\Delta$ \%} & \textbf{Recall@30 $\Delta$ \%} \\
\midrule
\textbf{Sampled Cross Entropy} & -19\% & -13\% \\
\textbf{Binary Cross Entropy} \cite{sasrec} & \textbf{-7\%} & \textbf{0\%} \\
\textbf{BPR} & -32\% & -11\% \\
\textbf{TOP1} & -73\% & -20\% \\
\hline
\end{tabular}
\end{minipage}
\hfill
\begin{minipage}{0.35\textwidth}
\centering
\caption{Average outfit age in the top-30 recommendations when different freshness strategies are used.}
\label{table:freshness}
\begin{tabular}{lc}
\toprule
\textbf{Algorithm} & \textbf{Freshness@30} \\
\midrule
\textbf{LTR} & 53 \\
\textbf{AFRA} & 51 \\
\textbf{ + age feature} & 40 \\
\textbf{ + age decay} & \textbf{20} \\
\bottomrule
\end{tabular}
\end{minipage}
\end{table*}

\subsubsection{Freshness} 

Recommender systems have bias towards older items since those usually have more interactions. Users, however, prefer fresh content, although not at the expense of relevance \cite{youtube}. We focus on freshness in the "Get the Look" feed use case, which is our main entry point for fashion inspirations. We measure freshness of recommendations as the average age of the top-30 recommended outfits, given in days. \Autoref{table:freshness} shows this metric for the top-30 recommendations. We can observe that the previous approach, LTR, and AFRA provide similar freshness. Moreover, we have conducted experiments on how this metric could be improved, and considered two strategies for adjusting for freshness. The first one is inspired by \cite{youtube}, where an item age feature is added during training which is set to 0 during inference to "de-bias" old outfits that have higher chance of being interacted with. The second one introduces a tuned age exponential decay, simulating "content aging", with a half-life of 3 weeks obtained by parameter tuning. This value is used to weight the ranking score produced by the model during inference, which can be seen as a re-ranking of the results. Exponential decay strategy proved to be particularly effective as it does not harm relevance and substantially increase fresh content among the top-$k$ recommendations, decreasing the average age from 51 to 20 days. On the other hand, the age feature strategy decreased the average age to 40 days.

\begin{table*}[t]
  \caption{A/B test results showing increase in user retention and engagement metrics of our transformer-based recommender system, compared against the existing algorithms for three use cases illustrated in \Autoref{fig:usecases}.}
\label{tab:reco}
\begin{tabular}{p{10em}cccccc}
\toprule
\multirow{2}{*}{\textbf{Customer segment}} & \multicolumn{2}{c}{\textbf{Get the Look}} & \multicolumn{2}{c}{\textbf{Style preview}} & \multicolumn{2}{c}{\textbf{Inspired by you}} \\
 & \textbf{Retention} & \textbf{Engagement} & \textbf{Retention} & \textbf{Engagement} & \textbf{Retention} & \textbf{Engagement} \\
\midrule
 \textbf{All customers} & \multirow{1}{*}{+28.5\%} & \multirow{1}{*}{+33.1\%} & \multirow{1}{*}{+23.9\%} & \multirow{1}{*}{+30.2\%} & \multirow{1}{*}{+130.5\%} & \multirow{1}{*}{+201.3\%} \\    
 \textbf{New customers \newline (cold-start)} & \multirow{2}{*}{+42.0\%}  & \multirow{2}{*}{+47.1\%}  & \multirow{2}{*}{+39.7\%}   & \multirow{2}{*}{+39.7\%}  & \multirow{2}{*}{+109.2\%}  & \multirow{2}{*}{+137.5\%} \\
 \textbf{Existing customers} & \multirow{1}{*}{+27.0\%} & \multirow{1}{*}{+32.0\%} & \multirow{1}{*}{+23.1\%} & \multirow{1}{*}{+29.6\%} & \multirow{1}{*}{+130.5\%} & \multirow{1}{*}{+201.9\%} \\
\bottomrule
\end{tabular}
\end{table*}

\subsubsection{Diversity}

Diversity is another aspect of recommender systems important to prevent filter bubbles that cause the customers to lose interest over time due to recommendations that are too similar \cite{relevancediversity}. We measure two types of diversity: \emph{inter-list diversity} (measuring the content diversity within a list of recommendations) and \emph{temporal-diversity} (measuring the difference in recommendations from one to the next visit). As a proxy for inter-list diversity we use the maximum consecutive sublist in the top-k created by the same creator (consecutive recommendations from the same creator are undesirable for our algorithm). AFRA and LTR had the highest diversity among all algorithms. For both algorithms this metric on average is less than 2.0, with AFRA outperforming LTR by up to 20\%. We define temporal diversity as a normalized set difference between the recommendation lists from two consecutive visits. Both algorithms perform similarly with temporal diversity roughly around 70\%. As a future work, we would like to introduce impression data to AFRA to improve temporal diversity by down-ranking items the user has already seen but not interacted with.

\subsubsection{Loss functions}

We employ other standard ranking functions based on negative sampling such as BPR \cite{bpr}, TOP1 \cite{rnns} and binary-cross entropy \cite{sasrec} to improve relevance and/or training speed. \Autoref{table:losses} shows the relative change in recall compared to standard categorical cross-entropy (softmax loss). We experiment with using 30 and 100 negative samples (without replacement). The main observation is that loss functions based on negative sampling are not very effective in our setting. The training speed improvements obtained are modest and always less than 2x. Hence, the decrease in relevance does not justify the improvements in training speed. One of the possible reasons for decreased relevance metrics could be the choice of negative samples not matching the background distribution well. Improving the negative sampling distribution as well as using hard negatives are among our future work directions.

\section{ONLINE RESULTS}
\label{online-evaluation}

In order to confirm the efficacy of our algorithm in real scenarios, we have performed A/B tests on three of our use cases: "Get the Look", "Style preview" and "Inspired by you" (\Autoref{fig:usecase-gtl};
 \Autoref{fig:usecase-style}, \Autoref{fig:usecase-inspired} respectively). Each A/B test was run for 3 to 4 weeks (until convergence). For both "Get the Look" and "Style preview", we compare AFRA-Batch against LTR in order to be consistent with the previous approaches that used daily updates of customer interaction data. Real-time session data is used only for new customers to address the cold-start problem. For the "Inspired by you" use case we use near real-time data and therefore compare AFRA-RT against the IB-CNN-kNN algorithm that uses near real-time data as well. \Autoref{tab:reco} summarizes the results from the A/B tests on the \emph{retention} and \emph{engagement} KPIs. We define retention as the share of users with multiple interactions within 7 days, and engagement as outfit interaction rate per user.

In summary, AFRA performs substantially better on all KPIs on all tested use cases. We can observe that on the "Get the Look" and "Style preview" use cases the improvements in retention range from 23\% up to 42\%. The improvement was even higher on the "Inspired by you" use case where it ranged from 109\% to 130\%. We observe a strong improvement on the engagement KPI as well that ranges from 30\% to 47\% for the first two use cases, reaching 201\% on "Inspired by you". On both KPIs, the improvements are stronger on the cold-start customers on the first two use cases, thanks to introducing in-session recommendations and contextual inputs in AFRA.

These strong results are consistent with our offline experiments. We believe this is because we train on diverse sources of interaction data from all products and premises which in turn helps to provide more personalized and relevant content and dampens feedback loops and selection biases \cite{netflixdl, criticialfashionxrecsys}.

\section{CONCLUSIONS}
\label{conclusions}

In this paper we have presented a reusable Transformer-based recommender system that is able to utilize different types of interactions with various fashion entities. We have shown our approach is able to model short term customer interests by providing session-based recommendations as well as take into consideration long-term user preferences and contextual information about the customer and the use case. We have demonstrated its effectiveness on different use cases, with extensive offline and online experiments that show that our approach substantially improves both customer retention and engagement.

Future work that we consider worth exploring is introducing impressions to AFRA to improve temporal diversity by organically down-ranking items the user has already seen but not interacted with. Furthermore, we would like to abstract the prediction head of our recommender system to allow the flexibility of re-ranking and hence the ability to apply AFRA on use cases with many millions of items. 

\bibliographystyle{ACM-Reference-Format}
\bibliography{references}

\end{document}